# Thin film superconducting quantum interferometer with ultralow inductance

S. I. Bondarenko and A. V. Krevsun

*B. I. Verkin Institute of Low-Temperature Physics and Engineering, National Academy of Sciences of Ukraine, pr. Nauki 47, Kharkov 61103, Ukraine*

E. V. Ilichev and U. Hubner

*Leibniz Institute of Photonic Technology, Albert-Einstein-Straße 9 (Beutenberg Campus), 07745 Jena, Germany*

V. P. Koverya and S. I. Link

*B. I. Verkin Institute of Low-Temperature Physics and Engineering, National Academy of Sciences of Ukraine, pr. Nauki 47, Kharkov 61103, Ukraine*

A simple method has been developed for manufacturing a thin film superconducting quantum interferometer (SQI) with ultralow inductance ($\sim 10^{-13}$ H). Current-voltage and voltage-field characteristics of the SQI are presented. The basic design equations are obtained and confirmed experimentally. The SQI has been used for the first time to determine the penetration depth of a magnetic field into a film of 50% In-50% Sn alloy.

## 1. Introduction

The experimental discovery of quantum interference in a doubly connected superconductor with two Josephson junctions in 1964 (Ref. 1) led to the development of various kinds of superconducting quantum interference devices (SQUIDs).[2] At present, superconducting magnets and SQUIDs are more widely used in science and technology than other superconducting devices. The major attraction of SQUIDs is their extremely high sensitivity to increments in an external magnetic field at low and infralow frequencies and the small size of their sensor elements. The sensor in a SQUID is a superconducting quantum interferometer (SQI). The SQI is most often configured as a miniature doubly connected superconducting circuit with one or two Josephson junctions.[2] Although SQIs have been built for many years, improvements in their design are ongoing and their areas of application are expanding. At present, thin film SQIs with two Josephson junctions are preferred. The thin film design makes it possible to use advances in modern nanotechnology for the components of SQIs in order to miniaturize them further. Micro- and nano-SQUIDs are needed for studies of the magnetic properties of organic and inorganic molecules, as well as for constructing modern SQUID microscopes with high spatial resolution.[3] In particular, the inductance $L_0$ of these kinds of nano-SQUIDs are now in the range of a few picohenries ($L_0 \approx 10^{-12}$ H). Further reductions in $L_0$ will make it possible to use Josephson junctions with substantially larger areas and critical currents, since the modulation depth $\delta i$ of the quantized current $i$ flowing in the SQI will be larger according to the equation[2]

$$\delta i = \Phi_0/(2L_0), \quad (1)$$

where $\Phi_0$ is the quantum of magnetic flux ($\Phi_0 = 2 \times 10^{-15}$ Wb). Increasing the size of the contacts reduces the effect of fluctuations in the critical current of the SQI on its characteristics, increases the noise resistance of the SQI, and makes it cheaper to manufacture the contacts.

Here the interference period of the change in the circulating current under the influence of an external magnetic field through the interferometer is given by

$$\Delta i = \Phi_0/L_0. \quad (2)$$

It should also be noted that the traditionally $L_0$ is taken to be the so-called magnetic inductance of the SQI.

## 2. The purpose of this work

The main purpose of this work has been to create a low-inductance SQI for a new superconducting interference structure that we have developed in the form of a doubly-connected superconductor (DSC) with an SQI. This structure consists of a high-inductance (circuit inductance greater $10^{-9}$ H) superconducting circuit with an asymmetric SQI with two Josephson junctions in its gap.[4–6] Up to now we have used a niobium-niobium compression point contact (CPC) for an SQI of this type. Figure 1 shows a conceptual drawing of a doubly-connected structure with a CPC and its equivalent circuit.

Estimates of the inductance of an SQI with this design yield about $10^{-13}$ H (0.1 pH). It is difficult to determine the size and critical currents for the individual microcontacts in this kind of SQI with the required accuracy. An improved theory of the processes in the new structure required a thin-film low-inductance two-contact SQI with sufficiently accurately determined sizes for its components and stable critical junction currents. The other purpose of this work was to develop a technology for manufacturing a low-inductance SQI that would be consistent with the current state of our experimental base and would be much simpler than those used in other laboratories.

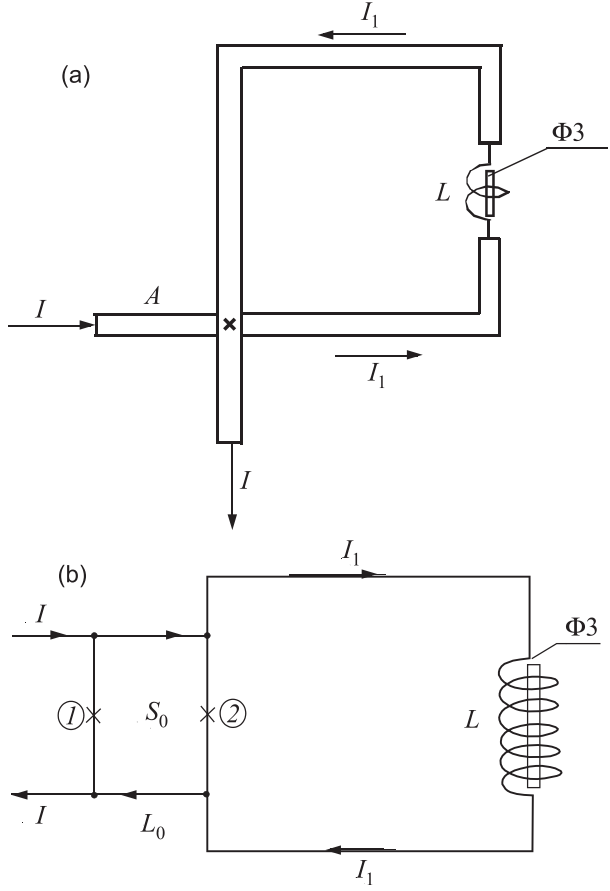

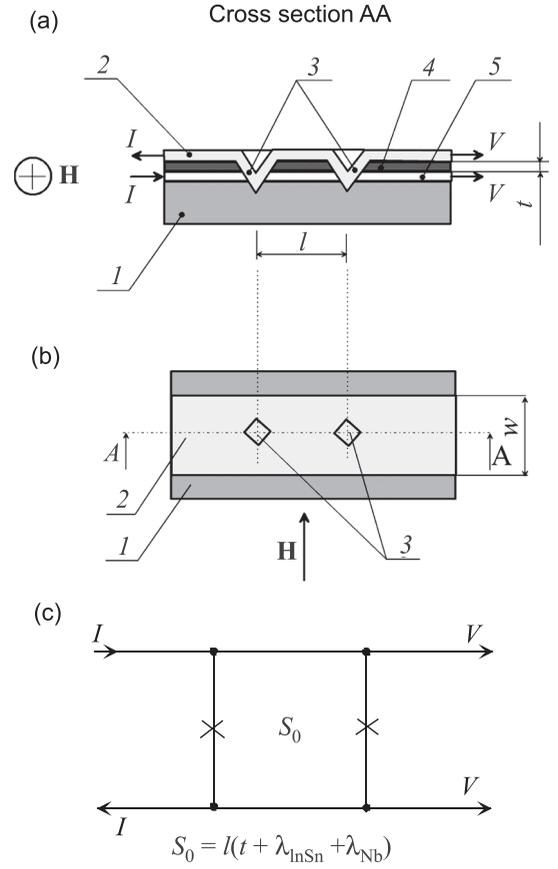

Fig. 1. (a) Conceptual diagram of a DSC with a CPC at the point of intersection of the wires $A$; $\Phi 3$ is the sensor of the ferroprobe magnetometer that measures the magnetic field of the current $I_1$; $L$ is the inductance of the DSC circuit; $I$ is the dc transport current. (b) Equivalent electrical circuit of a DSC with a CPC containing two Josephson junctions (1) and (2) with a quantization area $S_0$ and inductance $L_0$.

Fig. 2. (a) Transverse cross section of an SQI with two thin-film Josephson junctions (3) between a niobium film (5) and an InSn alloy film (2); (1) silicon substrate; (4) insulating niobium oxide with thickness $t$; $l$ is the distance between the junctions; I-I and V-V are the current and potential leads; $H$ is the measured magnetic field. (b) View of the surface of the SQI; w is the width of the InSn film; 3 denotes the imprints of the diamond pyramid at the locations where the InSn film is punctured to form the contacts. (c) Equivalent circuit of the SQI with two Josephson junctions (×) and a quantization area $S_0$.

### 3. Design and method for manufacturing the thin-film SQI

Figures 2(a) and 2(b), are schematic drawings in two projections of the asymmetric SQI developed here and Fig. 2(c) shows the electrical circuit with the current and potential leads.

The superconducting circuit of the SQI is formed by two thin films 2 and 5 with different widths and two contacts between them. The rectangular $1.5 \times 7$ mm niobium bottom film (5) with a thickness $t_5 = 100$ nm is deposited on a substrate of single crystal silicon (1) and electrolytically oxidized to a thickness $t_4 = 30$ nm. The narrow rectangular top plate of 50% Sn-50% In alloy with dimensions of $1.5 \times 7$ mm and thickness $t_2 \approx 200$ nm is deposited on the niobium oxide (4) of the lower film. The top and bottom plates are connected by two thin-film quasi-point Josephson junctions (3). The distance between the contacts ($l$) is 200 $\mu$m. The quantization area $S_0$ of the SQI is equal to

$$S_0 = l(t_4 + \lambda_2 + \lambda_5), \quad (3)$$

where $\lambda_2$ and $\lambda_5$ are the penetration depths of the magnetic field into the top and bottom films. The contacts were formed by mechanical puncturing of the films and the oxide layer using a standard PMT-3 diamond pyramid hardness measurement system. The imprint of the pyramid on the top film was about 10 $\mu$m. We assume that the microstructure of both of the contacts has the form shown in Fig. 3.

Because of the high plasticity of the InSn film, a metallic contact is formed between the InSn and Nb films around the perimeter of the pyramid during puncturing and entry of the pyramid into the substrate. The width of the contact is equal to the product of the perimeter and the thickness of the Nb film, and its length equals the thickness of the oxide (a typical value for compression junctions, i.e., close to atomic dimensions). Thus, the main specification for the metallic Josephson junctions is satisfied, i.e., its extent along the current flow should be on the order of or less than the coherence length $\xi$ of the metals in the contact (at 4.2 K, $\xi_{Nb} \approx 10$ nm and $\xi_{InSn} \approx 80$ nm).

### 4. Computational formulas

The experimentally observed parameter for a two-junction SQI is the period $\Delta H$ of the voltage $V$ of the SQI as a function of the external magnetic field $H$. For this it is necessary that the contacts of the SQI be in a resistive state determined by the dc current I through the SQI. According to the theory of SQI[3] with Eq. (3), $\Delta H$ should be given by

$$\Delta H = \frac{\Phi_0}{\mu_0 S_0} = \frac{\Phi_0}{\mu_0 l(t_4 + \lambda_2 + \lambda_5)}, \quad (4)$$

where $\mu_0 = 4\pi \times 10^{-7}$ H/m.

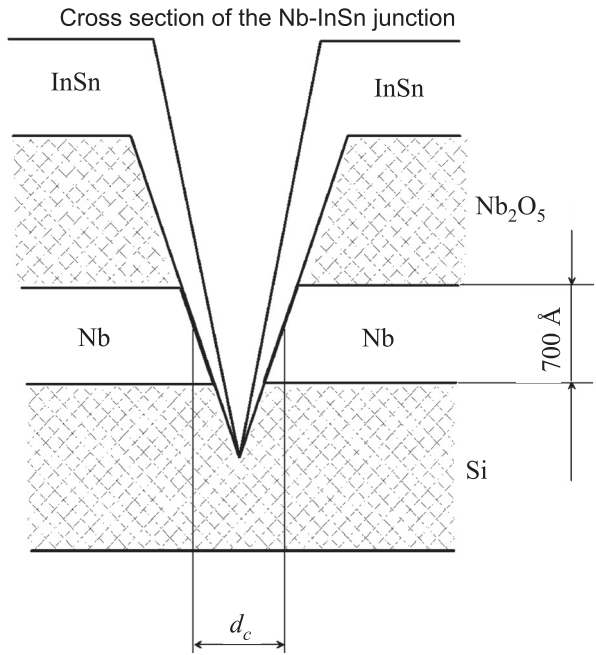

Fig. 3. Illustrating the cross section of one of the superconducting SQI junctions formed between the niobium and InSn films in a conical region with average diameter $d_c$ and a width of about 70 nm.

Now we determine the inductance of the interferometer. Measuring the low SQI inductances $L_0$ of a few pH or less is a fairly complicated task. Thus, here we limit ourselves to calculating the magnetic inductance of an SQI of the proposed type. The following ideas form the basis of this calculation. The magnetic flux $\Phi$ of an external magnetic field $H$ through the circuit of the SQI is fully compensated by the magnetic flux of the resulting circulation current $i$ as it increases initially ($\Phi < \Phi_0/2$) and then ($\Phi > \Phi_0/2$) penetrates into the loop in the form of a quantum of flux $\Phi_0$.[2] We consider only the region $\Phi < \Phi_0/2$, where $\Phi$ is equal to the flux of the field produced by the current $i$. Thus, the field $H$ and the field from the current $i$ are equal ($H = H_i$). The field $H_i$ is produced by the current flowing through the four parts of the loop: the segment in the narrow top film which produces a field $H_{it}$, the segment in the wide bottom film which produces a field $H_{ib}$, and the two segments with weak Josephson junctions, each of which produces a field $H_{il}$. Thus, the field $H_i$ can be written as the sum of the fields $H_{it}, H_{ib}, 2H_{il}$

$$H_i = H_{it} + H_{ib} + 2H_{il}. \quad (5)$$

The field $H_{it}$ is given by a formula[7] in SI units as

$$H_{it} = \frac{i}{10w}, \quad (6)$$

where $w$ is the width of the film. $H_{il}$ can be estimated as

$$H_{il} \approx \frac{i}{\pi l}. \quad (7)$$

The superconducting current $i$ along the wide bottom Nb film tends to spread to the edge. Because of the substantial (by 7.5 mm) distance of the edges of the Nb film from the quantizing loop of the SQI and the spreading of the current over its surface, we can neglect its contribution to the magnetic field $H_i$ compared to the other components. As a result, we obtain

$$H_i = H \approx \left(\frac{1}{10w} + \frac{2}{\pi l}\right)i. \quad (8)$$

Therefore, for a single period $\Delta H$ of the interference dependence $V(H)$ of the SQI, Eq. (8) implies that

$$\Delta H \approx \left(\frac{1}{10w} + \frac{2}{\pi l}\right)\Delta i. \quad (9)$$

From Eqs. (2) and (9) we obtain

$$L_0 \approx \Phi_0 \left(\frac{1}{10w} + \frac{2}{\pi l}\right)/\Delta H. \quad (10)$$

Given the experimental values of $\Delta H$, $l$, and $w$, Eq. (10) is a convenient way to calculate the inductance. For designing an SQI of this type, it is necessary to use another formula for the inductance $L_0$ which relates it only to the geometric dimensions of the required SQI. This formula is obtained by substituting $\Delta H$ from Eq. (4) in Eq. (10)

$$L_0 \approx \mu_0 \left(\frac{l}{10w} + 1\right)(t_4 + \lambda_2 + \lambda_5). \quad (11)$$

Thus, it can be seen from Eq. (11) that $L_0$ can be lowered by reducing the distance between the contacts and the thickness of the insulator between them, as well as by increasing the width of the top film. The reduction in the geometric inductances is bounded below by the Josephson inductance of the contacts ($L_J$) of the SQI, which is similar to the kinetic inductance of ordinary superconductors.[8] If $L_J$ is estimated using a formula from Ref. 8,

$$L_J = \frac{\Phi_0}{2\pi I_c}, \quad (12)$$

where $I_c$ is the critical current of the Josephson contact, then for $I_c = 10$ mA, we find $L_J \approx 3 \times 10^{-13}$ H. A more accurate calculation of $L_J$ will require knowledge of the exact geometry of both of the SQI junctions and proof of the applicability of Eq. (12) to junctions in the resistive state. These questions are currently open.

## 5. Experimental setup

In order to obtain the most information about the properties of SQI during a cryogenic experiment, four interferometers were mounted simultaneously on a single silicon substrate with a niobium film. The niobium films were obtained from the Institute of Photonic Technologies (Jena, FRG). After a layer of oxide of thickness 30 nm was electrochemically formed on niobium, four strips of indium-tin alloy with a thickness of about 100 nm were deposited through a mask in a vacuum system. Then the insulating oxide on the niobium was punctured at room temperature outside the vacuum chamber and a second strip of indium-tin alloy film with a thickness of about 100 nm was deposited. After the contacts were formed on the films and leads were soldered on, the sample was placed in a cryostat with liquid helium on a cryogenic mount. A photograph of one of the samples with the SQI is shown in Fig. 4.

The current-voltage and voltage-field characteristics of the interferometers were measured in liquid helium ($T = 4.2$ K). An external magnetic field $H$ was produced by a calibrated copper solenoid that encompassed the substrate with the interferometers. The measurement system consisted of regulated dc ($I^=$) and ac ($I^\sim$) current sources, a high-sensitivity dc microvoltmeter, an analog-to-digital converter, and a computer for displaying the current-voltage and voltage-field characteristics. A ferromagnetic shield was placed around the cryostat with the sample in order to reduce the effect of external parasitic electromagnetic fields.

## 6. Experimental results and discussion

The current-voltage characteristics of SQIs with a constant transport current had one feature in common. When the critical current on the SQI was reached, there was a jump in the voltage by 100 or more $\mu$V. The voltage after this was insensitive to the field $H$. A typical current-voltage characteristic of this type is shown in Fig. 5. The critical currents for most of the SQIs were in the range of 5–10 mA for normal resistances $R_N$ of 0.1–0.2 $\Omega$. Thus, there was no transition resistive region of the current-voltage characteristic between the purely superconducting and normal states of the SQI junctions. This made it impossible to obtain information on the voltage-field characteristic in the resistive state of the SQI.

One solution to this problem may be to improve the technique for fabricating the contacts in order to reduce the critical current at 4.2 K to a level of about 1 mA.[9] Another solution for already fabricated SQIs is to use a method we proposed earlier[10] for simultaneous passage of ac and dc transport currents through the junctions and to record the resulting dc voltage. As a result, the jump disappears from the recorded current-voltage characteristic and a resistive segment appears which is sensitive to the external magnetic field. Figure 6 shows a current-voltage characteristic obtained in this way for the same SQI and Fig. 7 shows the corresponding voltage-field characteristic.

A distinct periodic interference dependence $V(I_H)$, which is typical for dc SQI with two Josephson junctions, can be seen. $I_H$ is the current in the solenoid used to produce the field $H$. The period $\Delta I_H \approx 9$ mA is consistent to within 10% with the period in the magnetic field $\Delta H = 0.67$ Oe calculated using Eq. (4). $\lambda_2$ is taken to be the 70 nm thickness of the niobium film, since it was less than the penetration depth of the magnetic field into it.[11] Thus, it has been possible for the first time to measure the penetration depth $\lambda_5$ of the field into an InSn alloy film at $T = 4.2$ K: $\lambda_5 = (60 \pm 10\%)$ nm.

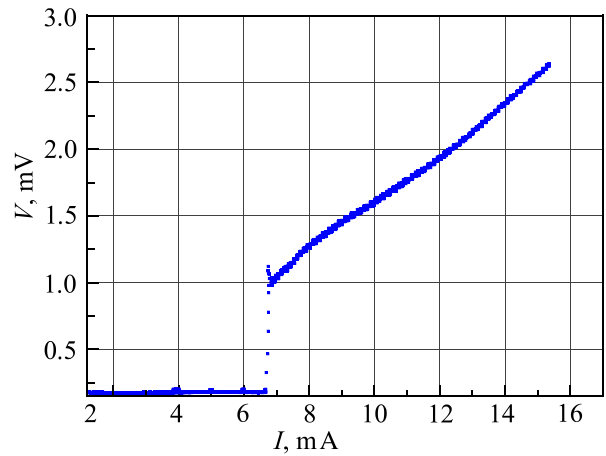

Fig. 5. Current-voltage characteristic of the SQI at $T = 4.2$ K. A jump in the voltage to 1.2 mV can be seen when the critical current of about $I_c^= = 6.5$ mA is reached.

The inductances calculated using Eqs. (10) and (11) are close to one another at about $10^{-13}$ H (i.e., 0.1 pH). The geometric inductance obtained here is close to the estimate for its minimum possible value for a critical current of $\sim$10 mA, i.e., to the Josephson inductance $L_J$ of the junctions in this SQI.

The technology described here for getting a low SQI geometric inductance $L_0$ can reduce it, in principle, by a factor of 20 by reducing the distance $l$ between the junctions. The reduction factor for the inductance is limited mainly by the optical resolution of the objective of the PMT-3 system and by the magnitude of $L_J$.

The question of comparing the inductance of the SQI with the modulation depth of the voltage cannot be answered now since the resistivity of SQI junctions of this type is determined by the combined action of dc and ac transport currents on them, while the standard theoretical assumptions only apply to the case of dc transport currents. It is known from experimental studies of the current-voltage characteristics of SQI junctions that adding an ac current both reduces the critical dc current of an SQI and greatly lowers the magnitude of the voltage modulation on the SQI by a magnetic field. This comparison becomes possible only after completion of the next stage in our development of low-inductance SQIs, which is aimed at significantly reducing their critical current for $T = 4.2$ K and getting current-voltage characteristics of the required form under our conditions using only a dc transport current.

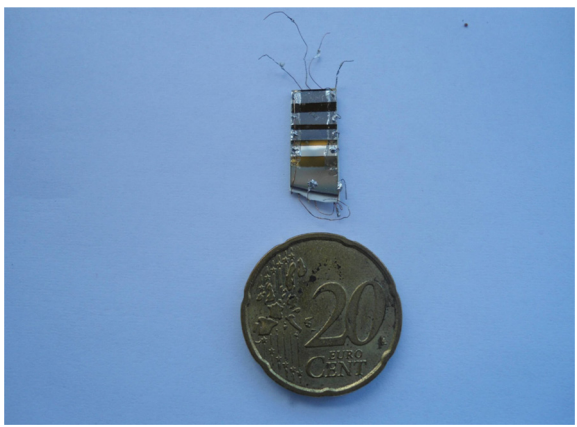

Fig. 4. A photograph of one of the samples with four SQIs, along with a 20 Euro cent coin for scale.

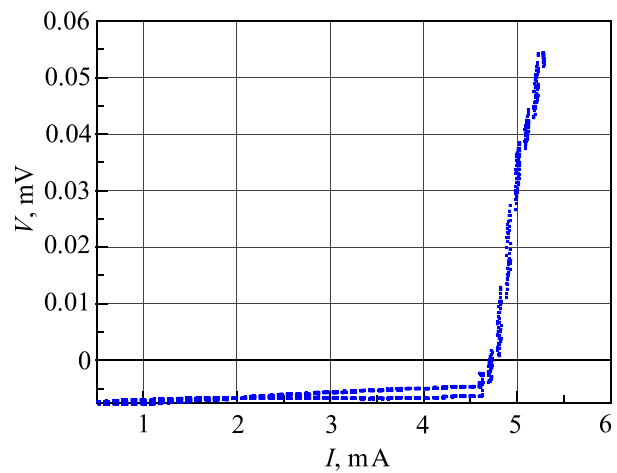

Fig. 6. Current voltage characteristic of the SQI at $T = 4.2$ K when a dc transport current is flowing and with a fixed ac current in the form of singly-polarized triangular pulses with a current amplitude of 2 mA at a frequency of $10^5$ Hz.

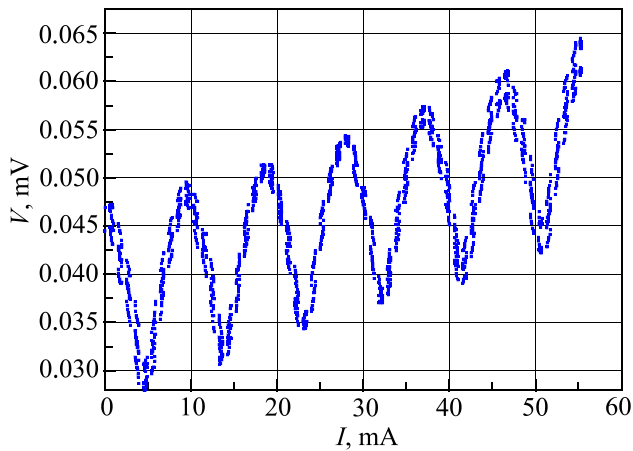

Fig. 7. Voltage-field dependence for the SQI with the current-voltage characteristic shown in Fig. 5. The amplitude of the quantum modulations in the voltage on the SQI is 20 $\mu$V with a period of the solenoid current $I_H$ used to produce the magnetic field $H$ of about 9 mA, which corresponds to $\Delta H$ = 0.67 Oe.

### 7. Conclusion

These experimental studies have led to the making of the first samples of thin-film superconducting quantum interferometers with a relatively simple design (in terms of technology) and an ultralow geometric (magnetic) inductance ($\sim 10^{-13}$ H).

Formulas for calculating the inductance of existing and designed interferometers have been obtained. In the first case, this is done on the basis of measuring the period $\Delta H$ of the interference dependence of the voltage $V$ on the interferometer as a function of external magnetic field $H$. In the second case, the formula for the inductance contains only the geometric dimensions of the interferometer components.

The experiments have shown that the computational formula for $\Delta H$ can be used to determine this quantity with high accuracy ($\pm 10\%$). In particular, this makes it possible to use measurements of $\Delta H$ as the simplest method for measuring the penetration depth of a magnetic field into the superconducting film in the loop for this kind of SQI. This method has been used for the first time in this paper to measure the magnetic field penetration depth into a film of 50% In-50% Sn alloy at 4.2 K: $\lambda_{InSn}$=60 $\pm$ 10% nm.

One of the priorities for further work with this type of SQI is to study the effect of a high shunt superconducting inductance ($10^{-6}$ H) on its characteristics. This will require a sufficiently high critical current (greater than the critical current for the Josephson junctions in SQIs) for the transition links between the SQI and the shunt inductance.

We thank A. S. Zaik and his colleagues for developing the analog-to-digital equipment to record the experimental data with a computer.

Email: bondarenko@ilt.kharkov.ua